\documentstyle[12pt,epsf]{article}
\begin{document}
\begin{titlepage}
\Large
\center{Depolarization of the cosmic microwave background}

\center{by a primordial magnetic field}

\center{and its effect upon temperature anisotropy} 

\vspace{0.5cm}
\normalsize
\center{Diego D. Harari,$^{\rm a,b,}$\footnote{Email 
address: harari@df.uba.ar}
 Justin D. Hayward$^{\rm a,c,}$\footnote{Email address: 
J.D.Hayward@damtp.cam.ac.uk}
and Matias Zaldarriaga$^{\rm d,}$\footnote{Email address:
matiasz@arcturus.mit.edu}} 
\vspace{0.3cm}
\center{ 
$^{\rm a}$ D.A.R.C., Observatoire de Paris - Meudon\\
5 Place Jules Janssen, 92195 Meudon, France}
\vspace{0.3cm}
\center{
$^{\rm b}$ Departamento de F{\'\i}sica, Facultad de Ciencias 
Exactas y Naturales\\ Universidad de Buenos Aires, Ciudad Universitaria - 
Pab. 1\\
1428 Buenos Aires, Argentina}
\vspace{0.3cm}
\center{
$^{\rm c}$ Department of Applied Mathematics and Theoretical Physics\\
University of Cambridge, Cambridge CB3 9EW, UK}
\vspace{0.3cm}
\center{
$^{\rm d}$ Department of Physics, MIT, Cambridge, MA 02139 USA}
\vspace{0.3cm}
\begin{abstract}  
We estimate the depolarizing effect of a primordial 
magnetic field upon the cosmic microwave background 
radiation due to differential Faraday rotation across the 
last scattering surface. The degree of linear polarization
of the CMB is significantly reduced 
at frequencies around and below 30 GHz
$(B_* /10^{-2}{\rm Gauss})^{1/2}$, where $B_*$
is the value of the primordial field at recombination. 
The depolarizing mechanism reduces the damping
of anisotropies due to photon diffusion 
on small angular scales.
The $l\approx 1000$ multipoles of the CMB temperature anisotropy 
correlation function in a standard cold dark matter cosmology
increase by up to 7.5\% at frequencies where depolarization 
is significant. 
    
\end{abstract} 
\end{titlepage}
\newpage

\section{Introduction} 

The cosmic microwave background radiation (CMB) is expected to have 
acquired a small degree of linear polarization  through Thomson
scattering \cite{Rees68}, which polarizes the radiation
if there is a quadrupole anisotropy in its  distribution 
function \cite{Chandrasekar50}. 
Typically, the CMB degree of linear polarization is expected to
be more than ten times smaller than the
relative temperature anisotropy on comparable angular scales, 
at least within a standard ionization history of the Universe.
The CMB has not yet been observed to be polarized, the 
upper limit on its degree of linear polarization on
large angular scales being
$P<6\times 10^{-5}$ \cite{Lubin83}. 
When measured, the CMB polarization will provide 
a wealth of information about the early Universe,
additional to that revealed by the CMB anisotropy.
 
The polarization properties of the CMB may prove particularly
valuable to either constrain or detect an hypothetical   
primordial magnetic field  \cite{Milaneschi85,Kosowsky96b}.
 A cosmological magnetic field 
could leave significant imprints upon the CMB polarization
through the effect of Faraday rotation.
After traversing a distance $L$ in a direction $\hat q$
within an homogeneous magnetic
field $\vec B$, linearly polarized radiation 
has its plane of polarization rotated an angle 
\begin{equation} 
\varphi={e^3n_ex_e\vec B\cdot\hat q \over 8\pi^2 m^2c^2}\lambda^2L\ .
\label{FR}
\end{equation}
$n_e$ is the total number-density of electrons and $x_e$ its ionized
fraction.
$\lambda$ is the
wavelength of the radiation, $m$ is the electron mass, 
and $c$ is the speed of light. We work in 
Heaviside-Lorentz electromagnetic units $(\alpha=e^2/4\pi\approx 1/137$
is the fine structure constant if we take $\hbar = c = 1$).

Faraday rotation of synchrotron emission by distant galaxies 
serves, for instance, to estimate the value of galactic and extragalactic
magnetic fields  \cite{Breviews}.
Faraday rotation acts also as a depolarizing mechanism.
If an extended source emits polarized radiation, the total
outcome may become significantly depolarized by a magnetic field,
after the radiation
emanating from points at different depths within the source 
experience different amounts of Faraday rotation. 
This process affects significantly the radio emission
of galaxies and quasars \cite{Burn66}.

In this paper we analyze the depolarizing effect exerted by a 
primordial magnetic field upon
the CMB across the last scattering surface.
We consider a Robertson-Walker universe
with scalar, energy-density fluctuations,
and assume a standard thermal history. 
We make use of an analytic approach
\cite{Zaldarriaga95},
based on a recent refinement and extension \cite{Hu95}
of the tight-coupling
approximation \cite{Peebles70},
that highlights the physical process
responsible for the CMB polarization
and its dependence upon various cosmological parameters, 
while still yielding reasonably accurate results.
The  polarization
of the CMB is proportional to
the width of the last scattering surface (LSS),
the interval of time during which most
of the CMB  photons that we observe today last-scattered off free
electrons. A primordial magnetic field could prevent the  polarization
from growing across the full width of the LSS. 
We shall see that the effect is controlled by the dimensionless
and time-independent parameter
\begin{equation}
F\equiv \frac 3 {2\pi e} {B\over\nu^2}
\approx
0.7( {B_*\over 10^{-3} {\rm Gauss}}\Big )\Big ({10 {\rm GHz}
\over \nu_0}\Big )^2\quad . \label{F}
\end{equation}
The coefficient $F$ represents the average Faraday rotation
(in radians) between Thomson scatterings \cite{Milaneschi85}. 
$\nu_0$ is the CMB frequency observed today.
$B_*=B(t_*)$ is the strength
of the primordial magnetic field at a redshift
$z_*=1000$, around the time of decoupling of matter and
radiation.
Current bounds  suggest that a 
magnetic field pervading cosmological  distances,
if it exists, 
should have a present strength below $B_0\approx 10^{-9}$ Gauss
\cite{Breviews}.
It is conceivable that the large scale magnetic
fields observed in galaxies and clusters have their
origin in a primordial field, and several
theoretical speculations exist 
about its possible origin \cite{Btheory}.
A primordial magnetic field 
is expected to scale as $B(t)=B(t_0)
a^2(t_0)/a^2(t)$, where $a(t)$ is the Robertson-Walker scale factor.
Thus, a primordial field with strength
$B_*=10^{-3}$ Gauss at recombination would have a present
strength
\begin{equation}
B_0=\frac {B_*}{(1+z_*)^2}\approx 10^{-9}\ {\rm Gauss}\ 
(\frac {B_*}{10^{-3}\ {\rm Gauss}\ }) \quad .
\end{equation}

A primordial magnetic field may significantly depolarize the
CMB right before its decoupling from matter. The effect is  sensitive
to the strength of the magnetic field at recombination,
not to its present strength. A  value of $B_*$ somewhat larger 
than $10^{-3}$ Gauss is not ruled out. Compatibility
with big-bang nucleosynthesis, for instance, places an upper bound that,
extrapolated to the time of recombination, is at most $B_*=0.1$
Gauss \cite{Bnuc}. Recent proposals for either detecting or constraining a
primordial field at recombination were suggested in \cite{Kosowsky96b,Adams96}.
In \cite{Kosowsky96b}, Faraday rotation of the CMB polarization
was analyzed in the limit of small rotation angles, concluding that 
a measurement of the effect could provide evidence for magnetic 
fields of order $B_*\approx 10^{-3}$ Gauss at recombination.
In \cite{Adams96}  the change in the photon-baryon
sound speed in the presence of a magnetic field of order 
$B_*=0.2$ Gauss was claimed to distort the structure
of the acoustic peaks in the CMB anisotropy power spectrum
at a level detectable by currently planned CMB experiments.

We shall entertain in our discussions 
the possibility that the strength of the primordial magnetic field 
at recombination be somewhat larger than $B_*=10^{-3}$ Gauss.
We will show that currently planned CMB experiments might be
sensitive to the effect of depolarization upon the temperature 
anisotropy power spectrum on small angular scales if
$B_*$ is around or larger than 0.01 Gauss, while experiments
at somewhat lower frequencies would be sensitive to primordial fields of
strength around $B_*\approx 10^{-3}$ Gauss.
 
The impact of depolarization upon anisotropy comes
about as a consequence of the polarization-dependence of Thomson 
scattering, which
feeds back polarization into anisotropy 
\cite{Zaldarriaga95,Hu95b,Kaiser83}. 
The dominant effect is a
reduction in the exponential damping due to photon
diffusion, which results in an increase of the
anisotropy at those frequencies for which depolarization
is significant. We shall perform an analytic estimate
of the effect, based on the tight coupling approximation. 
In order to make more quantitative and specific 
predictions about the impact and potential
measurability of the effect of depolarization upon 
temperature anisotropy, we shall also use   
a recently developed numerical code \cite{Seljak96}
to integrate the Boltzmann equations in a standard cold dark matter
model. We shall see that the temperature anisotropy
correlation function multipoles at $l\approx 1000$
increase by up to 7.5\% at frequencies where 
depolarization is significant.
We conclude that a primordial magnetic field of
strength around $10^{-2}$ Gauss
at recombination is worth
of membership in  the list of  multiple cosmological 
parameters that one may attempt to determine through CMB 
anisotropy measurements
on small angular scales \cite{Jungman96}.

The paper is structured as follows.
In section II, we write down and describe the radiative transfer equations 
for the total and polarized photon-distribution function
in the presence of a single Fourier mode of the scalar metric fluctuations. 
We include the term describing Faraday rotation by a primordial
magnetic field.  We solve these equations in the tight 
coupling approximation, and find the dependence of the
degree of polarization upon the frequency of the CMB photons
in the presence of a primordial magnetic field. 
In section III we discuss the effects of the depolarizing
mechanism upon the anisotropy of the CMB on small angular scales,
both analytically as well as numerically. We discuss
the possibility that the effect be detected by currently
planned CMB experiments. Section IV is the discussion and conclusion.

\section{Depolarization by a magnetic field}
\subsection{Boltzmann equations}

We begin by considering the radiative transfer equations for a single 
Fourier mode of the temperature and polarization
fluctuations in a Robertson-Walker spatially-flat
Universe with scalar (energy-density) metric fluctuations,
described in terms of the gauge-invariant gravitational potentials
$\Psi$ and $\Phi$. 
We follow the notation and formalism of Ref. \cite{Zaldarriaga95}.
The total temperature fluctuation is denoted by 
$\Delta_T$, while the fluctuation in the Stokes parameters
$Q$ and $U$ are denoted by $\Delta_Q$ and $\Delta_U$ respectively. 
The degree of linear polarization
is given by $\Delta_P=(\Delta_Q^2+\Delta_U^2)^{1/2}$.
All three quantities are expanded in Legendre
polynomials as
$\Delta_X=\sum_l (2l+1)\Delta_{X_l}P_l(\mu)$,
where $\mu=\cos\theta=\vec k\cdot\hat q/|\vec k|$ is the
cosine of the angle between the wave vector of a given Fourier mode 
$\vec k$,
and the direction of photon propagation $\hat q$. 
The evolution equations for the Fourier mode of wave vector $\vec k$
of the gauge-invariant temperature and polarization fluctuations
\cite{Zaldarriaga95,Hu95,Bond84,Kosowsky96a},
including the Faraday rotation effect of a primordial magnetic 
field \cite{Kosowsky96b}, read
\begin{equation}
\dot\Delta_T
+ik\mu(\Delta_T+\Psi)=-\dot\Phi -\dot\kappa
[ \Delta_T-\Delta_{T_0}-\mu V_b+\frac 1 2 P_2(\mu)S_P]\label{T}
\end{equation}
\begin{equation}
\dot\Delta_Q
+ik\mu\Delta_Q=-\dot\kappa [\Delta_Q-\frac 1 2(1- P_2(\mu)) S_P]
+2\omega_B\Delta_U\label{Q}
\end{equation}
\begin{equation}
\dot\Delta_U
+ik\mu\Delta_U=-\dot\kappa\Delta_U - 2\omega_B\Delta_Q\quad .\label{U}
\end{equation}
We have defined 
\begin{equation}
S_P\equiv -\Delta_{T_2}
-\Delta_{Q_2}+\Delta_{Q_0}
\label{SP}
\end{equation} 
which acts as 
the effective source term for the polarization. 
$V_b$ is the bulk velocity of the baryons, 
which verifies the continuity equation
\begin{equation}
\dot V_b=-{\dot a\over a}V_b-ik\Psi +{\dot\kappa\over R}
(3\Delta_{T_1}- V_b)\quad .
\label{VB}
\end{equation}
An overdot means derivative with respect to the conformal time
$\tau=\int dta_0/a$, with $a(t)$ the scale factor of the spatially flat
Robertson-Walker metric, and $a_0=a(t_0)$ its value at the present
time.  
$R\equiv 3\rho_b/4\rho_{\gamma}$ coincides with 
the scale factor $a(t)$ normalized to $3/4$ at
the time of equal baryon and radiation densities. 
$\dot\kappa=x_en_e\sigma_Ta/a_0$ is the Thomson scattering rate,
or differential optical
depth, with $n_e$ the electron number density, $x_e$ its ionized
fraction, and $\sigma_T$ the Thomson scattering cross-section. 
Finally, $\omega_B$ is the Faraday rotation rate 
\cite{Kosowsky96b}
\begin{equation}
\omega_B\equiv {d\varphi\over d\tau} = {e^3n_ex_e{\vec B}\cdot\hat q\over 
8\pi^2m^2\nu^2} {a\over a_0}\quad 
\end{equation}
If there were axial symmetry around $\vec k$ and no Faraday rotation,
one could always
choose a basis for the Stokes parameters such that $U=0$.
A magnetic field with arbitrary orientation breaks the axial symmetry,
and Faraday rotation mixes $Q$ and $U$. 

\subsection{\bf Tight coupling approximation}

We now solve the equations (\ref{T},\ref{Q},\ref{U}) 
in the tight - coupling
approximation, which amounts to an expansion in
powers of $k\tau_C$, where $\tau_C\equiv\dot\kappa^{-1}$
is the average conformal time between collisions.
 
At times earlier than decoupling, Thomson scattering is very
efficient, and the mean free path of the photons is very short.
The lowest order tight-coupling expression constitutes
in that case an excellent approximation. It implies that
the photon distribution function is isotropic in the
baryon's rest frame, and hence the polarization vanishes
\cite{Zaldarriaga95}. To first order in $k\tau_C$
there is a small quadrupole anisotropy, and thus a small
polarization. As decoupling of matter and radiation 
proceeds, the tight-coupling approximation breaks down.
Still, for wavelengths longer than the width of the
last scattering surface, it provides a very accurate
approximation to the exact result.

In the absence of a magnetic field ($\omega_B=0$), 
the tight - coupling solutions, to first order in $k\tau_C$, 
are such that \cite{Zaldarriaga95}  
\begin{equation}
\Delta_U=0\quad ;\quad \Delta_Q=\frac 34 S_P \sin^2\theta
\label{UQB=0}
\end{equation}
\begin{equation}
S_P=-\frac 52 \Delta_{T_2}=\frac 43 ik\tau_C\Delta_{T_1}
=-\frac 43\tau_C\dot\Delta_0\quad ,
\label{SPB=0}
\end{equation}
where we defined $\Delta_0\equiv\Delta_{T_0}+\Phi$.
Notice that $\Delta_{Q_0}=-5\Delta_{Q_2}=-\frac 54\Delta_{T_2}=
\frac 12 S_P$, while 
all multipoles with $l\geq 3$ vanish to first order in $k\tau_C$.
All quantities of
interest can be expressed, in the tight-coupling
approximation, in terms of $\Delta_0$, which in turn
verifies the equation of a forced and damped 
harmonic oscillator \cite{Hu95}
\begin{equation}
\ddot\Delta_0+
\Big [{\dot R\over 1+R} 
+\frac {16}{45}\frac {k^2\tau_C}{(1+R)}\Big ]
\dot\Delta_0+
{k^2\over 3(1+R)}\Delta_0=
\frac {k^2}{3(1+R)}[\Phi-(1+R)\Psi]
\quad ,
\label{T0B=0}
\end{equation}
where we have neglected $O(R^2)$ corrections.

Now consider the effect of the magnetic field ($\omega_B\neq 0$),
assumed spatially homogeneous over the scale of a perturbation 
with wave-vector $\vec k$. Faraday rotation breaks the axial 
symmetry around the direction of the wave-vector. 
The depolarizing effect of Faraday rotation depends not only
upon the angle between the magnetic field and the direction
in which the radiation propagates, but also upon the angle
between the magnetic field and the wavevector $\vec k$.
Nevertheless, we shall only be interested in the stochastic 
superposition of all Fourier modes of the density fluctuations, 
with a Gaussian spectrum that has no privileged direction. 
Average quantities thus depend only upon the angle between
the line of sight and the direction of the magnetic field,
but not upon the angle between the magnetic field and the
wavevector $\vec k$, which is integrated away.
For simplicity of the calculation, when computing the 
evolution of perturbations with wave-vector $\vec k$
we shall consider a magnetic field with no component 
perpendicular to 
$\vec k $. This choice  also satisfies the 
condition of axial symmetry around $\vec k$, under which  
eqs. (\ref{T},\ref{Q},\ref{U}) for $\Delta_T, \Delta_Q$
and $\Delta_U$ were derived.
We shall later use the result of this calculation
for the stochastic superposition of all Fourier modes
with arbitrary orientation relative to the magnetic
field. This simplification will result at most in an
underestimate of the net depolarizing effect,
since the case $\vec B\parallel\vec k$ is 
that for which depolarization is 
less effective,
the magnetic field being perpendicular
to the direction in which polarization is
maximum. 

To first order in $k\tau_C$ the tight-coupling solutions
in the presence of a homogeneous magnetic field $\vec B
\parallel\vec k$  are  such that:
\begin{equation}
\Delta_U=-F\cos\theta\Delta_Q\quad ;\quad
\Delta_Q=\frac 34 \frac{S_P \sin^2\theta}{(1+F^2\cos^2\theta)}
\label{UQTC}
\end{equation}
where we have defined the coefficient $F$ as     
\begin{equation}
F\cos\theta\equiv  2\omega_B\tau_C
\end{equation}
and so
\begin{equation}
F=
{e^3\over 4\pi^2m^2\sigma_T}\ {B\over\nu^2}
\approx 0.7
\Big ( {B_*\over 10^{-3} {\rm Gauss}}\Big )\Big ({10\ {\rm GHz}
\over \nu_0}\Big )^2\quad . \label{Fbis}
\end{equation}
The coefficient $F$ represents the average Faraday rotation 
between collisions, 
since $2\omega_B$ is the Faraday rotation rate 
and $\tau_C=\dot\kappa^{-1}$ is the photons mean free path 
(in conformal time units).
When calculating the evolution of each mode $\vec k$
we have assumed that the
strength of the primordial magnetic field scales as $B(t)=B(t_*)
a^2(t_*)/a^2(t)$, which is justified by flux conservation and because
the Universe behaves as a good conductor \cite{Breviews}. 
Since the frequency also
redshifts as $\nu=\nu_0 a(t_0)/a(t)$, the parameter $F$ is time-independent.
$\nu_0$ is the frequency of the CMB photons at present time, while
$B_*$ is the strength of the magnetic field at a redshift $z_*=1000$,
around recombination. Within a standard thermal history, with no early 
reionization, depolarization is only significant across the LSS, 
and it thus depends only upon the value of the primordial
magnetic field around the time of recombination.
Notice that Faraday rotation between collisions becomes
considerably large, paving the way to an efficient depolarizing
mechanism, at frequencies around and below $\nu_{\rm d}$ defined such that
\begin{equation}
F\equiv\Big (\frac {\nu_{\rm d}}{\nu_0}\Big )^2
\label{Fter}
\end{equation}
so that
\begin{equation}
\nu_{\rm d}\approx 8.4\ {\rm GHz}\ 
\Big(\frac {B_*}{10^{-3}{\rm Gauss}}\Big)^{1/2}
\approx 27\ {\rm GHz}\ 
\Big(\frac {B_*}{0.01\ {\rm Gauss}}\Big)^{1/2}\quad .
\label{nud}
\end{equation}

From eqs. (\ref{UQTC}) we can read the values of $\Delta_{Q_0}$
and $\Delta_{Q_2}$. They reduce to eqs. (\ref{UQB=0}) with
$O(F^2)$ corrections for small $F$, while they vanish as $F^{-1}$
for large $F$. We write them as:
\begin{equation}
\Delta_{Q_0}=\frac 12 d_0(F)S_P\quad ;\quad
\Delta_{Q_2}=-\frac 1{10} d_2(F) S_P\quad .
\label{d0d2}
\end{equation}
The coefficients $d_0,d_2$ are defined so that $d_i\approx 1+ O(F^2)$
for small $F$, while $d_i \rightarrow O(1/F)$ as $F\rightarrow\infty$,
and represent the effect of depolarization. They read:
\begin{equation}
d_0(F)=
\frac 32 \Big[\frac{\arctan(F)}{F}(1+\frac 1{F^2})-\frac 1{F^2}\Big ]
\end{equation}
\begin{equation}
d_2(F)=
\frac {15}{4}\Big[\frac{\arctan(F)}{F}(1+\frac 4{F^2}+\frac3{F^4})
-\frac 3{F^2}-\frac 3{F^4}\Big ]\quad .
\end{equation} 
In terms of the combination
\begin{equation}
d\equiv \frac 56(d_0+\frac{d_2}{5})=\frac{15}{8}
\Big[\frac{\arctan(F)}{F}(1+\frac 2{F^2}+\frac 1{F^4})-\frac 5{3F^2}
-\frac 1{F^4}\Big]
\label{d(F)} 
\end{equation}
and using the definition of $S_P$, we find the relation
\begin{equation}
\Delta_{T_2}=-S_P(1-\frac 35 d)
\label{T2TC}
\end{equation}
and from the equation for $\Delta_T$ in the
tight coupling limit we get
\begin{equation}
S_P=\frac 4{3(3-2d)} ik\tau_C\Delta_{T_1}
=-\frac 4{3(3-2d)}\tau_C\dot\Delta_0\quad .
\label{SPTC}
\end{equation}
Notice that $d\approx 1 - F^2/7$ if $F<<1$ while $d\rightarrow 
\frac {15}{16}\pi F^{-1}$ for large $F$.
We stress here again that in a general case the depolarizing coefficient 
$d$ depends upon the angle between $\vec k$ and
$\vec B$. The net anisotropy and polarization being the outcome of the
stochastic superposition of all Fourier modes of the density-fluctuations,
with a spectrum that has no privileged direction, the average 
depolarizing factor, after superposition of all wavevectors in arbitrary
orientations with respect to the magnetic field,
depends only upon $F$. The average depolarizing factor
might slightly differ from that calculated with  
$\vec k\parallel\vec B$, which at most
underestimates the average effect.

Equations (\ref{UQTC},\ref{T2TC},\ref{SPTC}) condense the main effects of  
a magnetic field upon polarization. 
When there is no magnetic field $(F=0,d=1)$
$\Delta_U=0$ and $\Delta_{Q}=-\frac {15}{8}\Delta_{T_2}\sin^2\theta$. A
 magnetic field generates $\Delta_U$, through Faraday rotation, 
 and reduces $\Delta_{Q}$. In the limit of very large $F$ 
(large Faraday rotation between collisions) the polarization vanishes. 
The quadrupole anisotropy $\Delta_{T_2}$ is also reduced by the
depolarizing effect of the
magnetic field, by a factor 5/6 in the large $F$ limit, because 
of the feedback of $\Delta_Q$ upon the anisotropy or, in
other words, because of the polarization dependence of
Thomson scattering. The dipole $\Delta_{T_1}$ and monopole
$\Delta_{T_0}$ are affected by the magnetic field only
through its incidence upon the damping mechanism due to photon
diffusion for small wavelengths, that we shall discuss in detail 
in section III. Indeed, the equation for $\Delta_0=\Delta_{T_0}+\Phi$,
neglecting $O(R^2)$ contributions, now reads:
\begin{equation}
\ddot\Delta_0+
\Big[{\dot R\over 1+R} 
+\frac {16}{90}\frac{(5-3d)}{(3-2d)}\frac {k^2\tau_C}{(1+R)}\Big]
\dot\Delta_0+
{k^2\over 3(1+R)}\Delta_0=
\frac {k^2}{3(1+R)}[\Phi-(1+R)\Psi]
\quad .
\label{T0TC}
\end{equation}
The damping term is reduced by a factor 5/6  at frequencies such that
$d<<1$, for which depolarization is significant.

We have assumed that the magnetic field is spatially homogeneous.
We can expect corrections to our result if the field is inhomogeneous
over scales smaller than $\tau_C$ at any given time around decoupling.
Indeed, if the field reverses its direction $N$ times along
a photon path during a time $\tau_C$, Faraday rotation will
not accumulate as assumed above. In that case depolarization
would start to be significant only at those frequencies such that
Faraday rotation is large over the scale on which 
the magnetic field reverses its direction. The  frequencies
at which depolarization starts to be significant would thus
be reduced by a factor $1/\sqrt N$.  

\subsection{Frequency-dependence of the degree of polarization}

The anisotropy and polarization observed at present time can be evaluated
using the formal solutions of eqs. (\ref{T},\ref{Q},\ref{U})
\begin{equation}
\begin{array}{ll}
\Delta_T(\tau_0)=&
\int_0^{\tau_0} d\tau e^{ik\mu(\tau-\tau_0)}g(\tau)
[\Delta_{T_0}(\tau)+\mu V_b(\tau) -\frac 1 2 P_2(\mu)S_P(\tau)]\cr
\ &\ \cr
\ &\quad +\int_0^{\tau_0} d\tau e^{ik\mu(\tau-\tau_0)}
e^{-\kappa(\tau_0,\tau)}(\dot\Psi  - \dot\Phi)
\end{array}
\label{TFormal}
\end{equation}
\begin{equation}
\Delta_Q(\tau_0)=\int_0^{\tau_0} d\tau e^{ik\mu(\tau-\tau_0)}g(\tau)
\{\frac 1 2 [1-P_2(\mu)]S_P(\tau)+F\Delta_U(\tau)\}
\label{QFormal}
\end{equation}
\begin{equation}
\Delta_U(\tau_0)=-\int_0^{\tau_0} d\tau e^{ik\mu(\tau-\tau_0)}g(\tau)
F\Delta_Q (\tau)
\label{UFormal}
\end{equation}
where 
\begin{equation}
g(\tau)\equiv\dot\kappa e^{-\kappa(\tau_0,\tau)}
\end{equation}
is the visibility function. It represents the probability that a photon
observed at $\tau_0$ last-scattered within $d\tau$ of a given $\tau$.
For a standard thermal history, with  no significant early reionization after 
recombination, $g(z)$ is well approximated by a Gaussian centered at a 
redshift of about $z\approx 1000$ and width $\Delta z\approx 80$
\cite{Jones85}. In conformal time, we shall denote the center and
width of the Gaussian which approximately describes
the process of decoupling by $\tau_D$ and $\Delta\tau_D$
respectively.
 
The visibility function being strongly peaked around the time of
decoupling, the first integral  in 
eq. (\ref{TFormal}) for the anisotropy
is well approximated, at least for wavelengths longer than the
width of the last scattering surface, by its instantaneous recombination
limit. In that case it reduces to the tight-coupling expression of
its integrand evaluated at time
$\tau=\tau_D$ \cite{Hu95}. 
This first integral is dominated by its first two terms,
proportional to the monopole $\Delta_{T_0}$ and the
baryon velocity $V_b$ (in turn proportional to $\Delta_{T_1}$)
respectively. 
The quadrupole term $S_P$ gives a negligible contribution
for long wavelengths, but becomes relatively significant
on small scales. The second integral
in eq. (\ref{TFormal}) corresponds to the anisotropies 
induced by time-dependent potentials after the time of last-scattering. 

Eqs. (\ref{QFormal},\ref{UFormal}) for the polarization
can be approximated replacing the integrand by its tight-coupling 
expression. Then
\begin{equation}
\Delta_Q(\tau_0)=\frac 34 
\frac {\sin^2\theta}{(1+F^2\cos^2\theta)}
\int_0^{\tau_0} d\tau e^{ik\cos\theta(\tau-\tau_0)}g(\tau)
S_P(\tau)
\label{QF2}
\end{equation}
\begin{equation}
\Delta_U(\tau_0)=-F\cos\theta\Delta_Q(\tau_0)
\label{UF2}
\end{equation}
while the total polarization, $\Delta_P=(\Delta_Q^2 +\Delta_U^2)^{1/2}$
reads
\begin{equation}
\Delta_P(\tau_0)=\sqrt{1+F^2\cos^2\theta}\Delta_Q(\tau_0)
\label{PF}
\end{equation}
Evaluation of the time integral in eq. (\ref{QF2})
requires a more 
detailed knowledge of the time-dependence of the integrand
than in the case of the anisotropy. 
Indeed, the 
tight-coupling expression (\ref{SPTC}) for the quadrupole
term $S_P$ being proportional to the mean free path 
$\tau_C$, which varies rapidly during decoupling,
the instantaneous recombination approximation becomes
inappropriate. The induced polarization is, indeed, 
proportional to the width of the last scattering surface.
Adapting the method of \cite{Zaldarriaga95} 
to include also the effect of the primordial magnetic field,
we write down the equation satisfied by $S_P$ 
when all other quantities are already approximated
by their first-order tight-coupling expressions:
\begin{equation}
\dot S_P+\frac 3 {10}(3-2d) \dot\kappa S_P=\frac 2 5 ik\Delta_{T_1}
\label{SP2}
\end{equation}
Neglect of $\dot S_P$ returns  the 
tight-coupling result of eq. (\ref{SPTC}).
Instead, the formal solution to equation (\ref{SP2})
\begin{equation}
S_P(\tau)={2\over 5} ik\int_0^{\tau'} d\tau'\Delta_{T_1}e^{-{3\over 10}
\kappa(\tau,\tau')(3-2d)}
\end{equation}
tracks down the time-dependence of $S_P$ through the decoupling
process with better accuracy. 

For wavelengths longer than the width of the LSS we can neglect the time 
variation of $\Delta_{T_1}$ and that of $e^{ik\cos\theta(\tau-\tau_0)}$
around decoupling. We also 
approximate the visibility function by a Gaussian,
which justifies the approximation 
$\dot\kappa(\tau_0,\tau)\approx{-\kappa(\tau_0,\tau)\over \Delta\tau_D}$
\cite{Polnarev85}.
Then
\begin{equation}
S_P(\tau)\approx\frac 2 5 ik\Delta_{T_1}(\tau_D)\Delta\tau_D e^{\frac 3 {10}
\kappa(\tau_0,\tau)(3-2d)}
\int_1^{\infty} \frac {dx}{x}
 e^{-\frac 3 {10}x\kappa (3-2d)}\quad ,
\label{SPter}
\end{equation}
where the integration variable has been changed to $x={\kappa(\tau_0,\tau)
\over \kappa(\tau_0,\tau')}$.
Thus, within these approximations,
\begin{equation}
\begin{array}{ll}
\int_0^{\tau_0} d\tau g(\tau) S_P(\tau)
&=-{2\over 5} ik\Delta_{T_1}(\tau_D)
\Delta\tau_D\int_0^{\infty}d\kappa 
e^{-\frac{1+6d}{10}\kappa}{\rm Ei}(-\frac{3}{10}(3-2d)\kappa)\cr
\ &\ \cr
\ &=\frac{4}{1+6d}ik\Delta_{T_1}(\tau_D)\Delta\tau_D
[\ln (\frac{10}{3})-\ln(3-2d)]\cr
\end{array}
\end{equation} 

Finally, the total polarization induced at an angle $\theta$ 
with respect to the wavector $\vec k$, reads 
\begin{equation}
\Delta_P(\tau_0)=\frac {3}{(1+6d)}[\ln(\frac{10}{3})-\ln(3-2d)]
\frac{\sin^2\theta e^{ik\cos\theta(\tau_D-\tau_0)}}
{\sqrt{1+F^2\cos^2\theta}}ik\Delta_{T_1}(\tau_D)\Delta\tau_D
\quad .
\label{PF0}
\end{equation}
It can also be written as follows, 
in terms of the polarization that would be induced if 
there were no magnetic field (or equivalently, in terms
of the polarization at frequencies large enough such that the depolarizing
effect is negligible): 
\begin{equation}
\Delta_P(\theta,F)=D(\theta,F)\Delta_P(B=0)
\end{equation}
where we have defined the depolarizing factor as
\begin{equation}
D(\theta,F)= \frac {1}{\sqrt{1+F^2\cos^2\theta}}
f(F)
\label{D(F)}
\end{equation}
with
\begin{equation}
f(F)= \frac{7}{1+6d}\Big[ 1-\frac{\ln(3-2d)}{\ln(10/3)} \Big]
\label{f(F)}
\end{equation} 
Eq. (\ref{D(F)}), together with the defining eqs. (\ref{Fbis},\ref{Fter})
and (\ref{d(F)})
for $F$ and $d$, 
summarize the main result of this section. 
Notice that
\begin{equation}
f\rightarrow 1 \quad {\rm as}\quad F\rightarrow 0
\quad ;\quad
f\rightarrow 0.61 \quad {\rm as}\quad F\rightarrow\infty
\end{equation}

The polarization observed at present times depends 
upon the angle between the line of sight and the orientation
of the magnetic field at the time of decoupling. There will be
no depolarization if the magnetic field is perpendicular
to the line of sight. The magnetic field is likely to change orientation
randomly over scales longer than the Hubble radius at the time of
decoupling, which subtends an angle of order one degree in the sky,
so that after averaging over many regions separated by more than a few
degrees, we can always expect a net average depolarizing effect.
To roughly estimate its order of magnitude we could assume 
an average component of 
$\vec B$ parallel to the observation
direction of order $B/\sqrt{2}$ and 
define an 
average $\bar D$ as
\begin{equation}
\bar D= \frac{1}{\sqrt{1+F^2/2}} f(F) \quad .
\label{Daverage}
\end{equation}
Fig. 1 displays
the depolarizing factor $\bar D$ as a function of the CMB 
frequency $\nu_0$.
We have plotted it for three different values of the magnetic
field $B_*$  to help visualize the relevant frequency
range, but notice that
since depolarization depends only upon $F=(\nu_{\rm d}/\nu_0)^2$, the plot
for an arbitrary  value of $B_*$ is identical to 
that corresponding to another value of the magnetic field after 
an appropriate scaling of the frequency units, proportional to the square
root of the magnetic field.

At low frequencies, those for which the effect
is large, the average depolarizing factor scales as 
\begin{equation}
\bar D\approx 0.6\frac{\sqrt{2}}{F}\approx 0.85
\Big( \frac{\nu_0}{\nu_{\rm d}}\Big)^2
\quad {\rm if}\quad \nu_0<<\nu_{\rm d}\quad .
\label{Dlow}
\end{equation}
At comparatively large frequencies instead
\begin{equation}
\bar D\approx 1 - 0.36 F^2= 1-0.36 
\Big(\frac{\nu_{\rm d}}{\nu_0}\Big)^4
\quad {\rm if}\quad \nu_0>>\nu_{\rm d}\quad .
\label{Dlarge}
\end{equation}

\section{Effects upon the anisotropy}

Depolarization by a primordial magnetic field
has significant and potentially measurable 
effects upon the anisotropy of the CMB on small angular scales.
Indeed, the polarization properties of the CMB feed back into
its anisotropy, as evidenced 
in eq. (\ref{T}), due to the polarization
dependence of Thomson scattering. The dominant effect of polarization
upon anisotropy derives from its impact upon
the photon diffusion length\cite{Zaldarriaga95,Hu95b,Kaiser83}, 
which damps anisotropies on small angular scales 
\cite{Hu95,Silk68,Peebles80}. It was shown in \cite{Hu95b},
through numerical integration of the Boltzmann equations,
that neglect of the polarization properties of the CMB leads to
an overestimate of its anisotropy on small angular scales as large
as 10\%. We thus expect depolarization by a primordial magnetic
field to introduce a significant frequency-dependent distortion 
of the CMB anisotropy power spectrum.  
Notice that a different (frequency-independent) 
distortion of the CMB anisotropy power spectrum
by a primordial magnetic field, due to its impact upon
the photon-baryon fluid sound speed,
was recently discussed in \cite{Adams96}.

\subsection{Reduced diffusion-damping}

Photon diffusion damps anisotropies on
small angular scales \cite{Hu95,Silk68,Peebles80}.
The effect is described by the term proportional to
$k^2\tau_C\dot\Delta_0$ in eq. (\ref{T0TC}).
The photon diffusion length depends upon the
degree of polarization of the CMB \cite{Zaldarriaga95,Hu95b,Kaiser83}.
Thus, the photon-diffusion length is different
at frequencies where the depolarizing effect is significant.

The damping of anisotropies on small angular scales
due to photon diffusion can be found, now including the
full $R$-dependence, by solving the
tight-coupling equations to second order, 
assuming solutions of the form 
\begin{equation}
\Delta_{X}(\tau)=\Delta_X e^{i\omega\tau}
\end{equation}
for $X=T$, $Q$, and $U$, and similarly for the baryon velocity $V_b$.
One then finds that 
\begin{equation}
\omega = \frac {k}{\sqrt{3(1+R)}} + i \gamma
\end{equation}
with the photon-diffusion damping length-scale determined by
\begin{equation}
\gamma (d)\equiv \frac{k^2}{k_D^{2}}= 
\frac {k^2\tau_C}{6(1+R)} \Big(\frac {8}{15}
\frac {(5-3d)}{(3-2d)} + \frac {R^2}{1+R}\Big)
\end{equation}
The depolarizing effect of the magnetic field reduces
the viscous damping of a\-ni\-so\-tro\-pies. 
In the case of small $R$, such that $R^2$
terms can be neglected, the damping factor $\gamma$
is smaller by a factor 5/6 at those frequencies for which the
depolarizing effect is large.

We make now an analytic estimate of the effect of the
frequency-dependence of the photon diffusion length,
in the presence of a primordial magnetic field, upon the
CMB anisotropy power spectrum. The temperature anisotropy
correlation function is typically expanded in Legendre
polynomials as 
\begin{equation}
C(\theta)=<\Delta_T({\hat n}_1)\Delta_T({\hat n}_2)>_{{\hat n}_1\cdot
{\hat n}_2=\cos\theta} = \frac 1 {4\pi}
\sum_{l=0}^\infty (2l+1)C_lP_l(\cos\theta)\quad .
\end{equation} 
The multipole coefficients of the anisotropy power spectrum are 
given by
\begin{equation}
C_l=(4\pi)^2 \int k^2 dk P(k)|\Delta_{T_l}(k,\tau_0)|^2
\end{equation}
with $P(k)$ the power-spectrum of the scalar
fluctuations, assumed  scale-invariant in the sCDM model.
The largest contribution to a given multipole $C_l$
comes from those wavelenghts such that $l=k(\tau_0-\tau_D)$,
where $\tau_0$ is the conformal time at 
present and $\tau_D$ the conformal time at decoupling.
The average damping factor due to photon diffusion upon the $C_l$'s 
is given by an integral of $e^{-2\gamma}$ times the visibility function
across the last scattering surface \cite{Hu95}. It depends upon cosmological
parameters, notably $R$, and upon  the recombination history.
Approximately, and for a standard cold dark matter model,
we can take $2\gamma (d=1)\approx (l/1500)^2$.   
The relative  change in the $C_l$'s
due to the change in the photon-diffusion length,
as we move down from frequencies where depolarization is 
insignificant ($d=1$) to lower frequencies ($d<<1$),
is then given by
\begin{equation}
\Delta C_l=\frac{C_l(d)}{C_l(d=1)}-1\approx \exp{\Big (
\frac{(l/1500)^2(1-d)}{(6-4d)} \Big )}-1
\quad . \label{Delta}
\end{equation}
In figure 2 we plot $\Delta C_l$ (expressed as a percentage) at $l=1000$
as a function of frequency, for three different values of 
the magnetic field $B_*=$ 0.001, 0.01 and 0.1 Gauss.
Once again, the graph for an arbitrary value of $B_*$ can be read from 
any of these with an appropriate scaling of the frequency units. 
We have chosen to display the effect at $l=1000$, 
that will be accessible by the recently funded
CMB satellite experiments, MAP \cite{MAP} and COBRAS/SAMBA \cite{CS}.

\subsection{Reduced quadrupole contribution}

The depolarizing effect of a primordial magnetic field
also changes the strength of the quadrupole term $S_P$
around decoupling, and thus its incidence upon the
anisotropy of the CMB on small angular scales.
Indeed, the quadrupole anisotropy and the polarization of the
CMB at the time of recombination contribute to the
presently observed anisotropy
through the following term of eq. (\ref{TFormal})
\begin{equation}
\Delta_{S_P}(\tau_0)\equiv
-\frac 12 P_2(\cos\theta)\int_0^{\tau_0} d\tau e^{ik\cos\theta(\tau-\tau_0)}
g(\tau)S_P(\tau)\quad .
\end{equation}
This term is negligible for long wavelengths, those that
dominate the lowest multipoles of the present anisotropy,
but becomes non-negligible on small angular scales
(large multipoles).
Indeed, in the tight coupling approximation
$S_P\propto \tau_C\dot\Delta_0$ and thus, barring 
a very strong time-dependence of the scalar potential,
the contribution of $S_P$ is well below
that of the monopole term, except for small wavelengths.

The depolarizing effect of a magnetic field
modifies the value of $S_P$ around decoupling, compared to what 
it would have been if there were no magnetic field, and then
\begin{equation}
\Delta_{S_P}(F)=  f (F) \Delta_{S_P}(B=0)
\end{equation}
with $f$ as defined in eq. (\ref{f(F)}).
Recall that $f\approx 1$ if $\nu >>\nu_{\rm d}$ while
$f\approx 0.6$ if $\nu<<\nu_{\rm d}$. Thus, at frequencies such that
the depolarizing effect of the magnetic field is significant,
the partial contribution of the quadrupole term $S_P$ to the 
total anisotropy is reduced by a factor 0.6 compared
to that at frequencies where depolarization is unimportant.
On small angular scales this could
represent a decrease of the anisotropy by a few per cent.
The effect is opposite to that of the change in diffusion
damping, but is likely to be less significant on small
angular scales.

\subsection{Numerical estimate of the effect upon the anisotropy}

In order to accurately ascertain    
the net effect of the depolarizing mechanism upon the 
CMB anisotropy and to make definite quantitative predictions within a
standard cosmological model, we turn now to the numerical integration of 
the Boltzmann equations (\ref{T},\ref{Q},\ref{U}). 
We use the recently developed code CMBFAST\cite{Seljak96}, 
that integrates  the sources
over the photon past light cone. Its starting point are the
formal solutions (\ref{TFormal},\ref{QFormal},\ref{UFormal}), 
where the geometrical and dynamical contributions are
separately handled to improve efficiency.  
As in our analytic estimates, when computing the evolution
of each Fourier mode
we introduce in the code the Faraday
rotation term with the angular dependence
corresponding to the case where $\vec B$ has no component
perpendicular to  $\vec k$.

Figures 3 and 4 summarize the numerical calculation of
the effect of depolarization upon temperature anisotropy, in 
a standard cold dark matter model (sCDM). 

The quantity plotted in Fig. 3 is $l(l+1)C_l$,
for the sCDM model without a magnetic field and
with a magnetic field and at frequencies such that
$F=1,4,9$, corresponding to $\nu_0=\nu_{\rm d},\nu_{\rm d}/2$ and
$\nu_{\rm d}/3$ respectively. 
Fig. 3 clearly shows that the CMB anisotropy 
on small angular scales increases
at frequencies where depolarization is significant.
This result indicates that the reduction in diffusion
damping due to depolarization is the dominant 
effect among the two opposite effects discussed 
in the previous  sections. 

Fig. 4 displays the same results but expressed 
in terms of $\Delta C_l$, the percentual
increase in  $C_l$ relative to the case without magnetic field.
The monotonic curves in the same figure, included for comparison
purposes, correspond to the analytic
estimate of the effect of reduced diffusion damping, eq.(\ref{Delta}).
As expected, the effect is larger on
smaller angular scales (larger $l$). The numerical result approximately
follows the analytic estimate of the effect of the reduction
in diffusion damping.
The total effect, however, does not
increase monotonically with $l$. This can be understood as a
consequence of the nature of the subdominant 
quadrupole contribution $S_P$, which oscillates out of
phase with the  $C_l$'s \cite{Zaldarriaga95} 
(remember that $S_P\propto \dot\Delta_0$), and is 
reduced by depolarization through a factor $f(F)$.

It is also clear from Fig. 4 that the analytic result
for the change in diffusion damping due to depolarization
overestimates the total effect at high l. This is because
the actual damping in the $C_l$ spectra has two contributions,
one from Silk damping and the other due to cancellations
in the integral across the last scattering surface produced
by the oscillations in the exponential and sources in
equation (\ref{TFormal}). Only  Silk damping is reduced
by the magnetic field, and that is why equation (\ref{Delta})
slightly overestimates the net effect. 

The analytic and numeric calculations are in very good agreement 
around $l\approx 1000$. 
The frequency-dependence of $\Delta C_l$ at $l=1000$
plotted in Fig. 2 fits very well the analogous result
after the full numerical integration of the Boltzmann equations.

We conclude that the depolarizing effect of the magnetic
field results in an increase of the anisotropy correlation
function multipoles of up to 7.5\%  (for sufficiently low frequencies)
on small angular
scales ($l\approx 1000$),
those that will be accessible
by future CMB satellite experiments such as MAP and COBRAS/SAMBA.
The frequencies at which the effect is significant, however,
depend on the strength and coherence length of
the primordial magnetic field at the time of recombination.
 
Depolarization depending upon frequency, the effect might
be difficult to separate from foreground contamination.
The relative change of the $C_l$'s at $l=1000$ 
is larger than 2\% on frequencies below 30 GHz (accessible
to the first two channels in MAP), 
if $B_*=0.02$ Gauss or larger.
The first two channels in COBRAS/SAMBA being at 31.5
and 53 GHz, the signal would reach a 2\% level within
this range only if $B_*$ is around or larger than
0.1 Gauss. However,
COBRAS/SAMBA might reach out to larger values of $l$,
where $\Delta C_l$
is larger, and might thus have a 
sensitivity to the depolarizing effect of $B_*$ 
comparable to MAP. In any case, both experiments
will be sensitive to a magnetic field around $B_*$=0.1 Gauss,
and would thus  at least be able to place a direct constraint
on $B_*$ comparable or better than the one obtained from 
extrapolation of the nucleosynthesis bound.

Experiments searching CMB anisotropy 
and polarization at smaller frequencies,
which  currently operate down to 5 GHz \cite{Partridge,JB},
may play a significant role to detect the depolarizing effect of
a primordial magnetic field.

\section{Conclusion}

The CMB is expected to have a small degree of
linear polarization. Several estimates  
were made for the predicted polarization, both in the context
of anisotropic cosmological models
\cite{Rees68,Pa},
as well as in isotropic and homogeneous cosmologies
perturbed with either energy-density fluctuations
or gravitational waves \cite{Bond84,Polnarev85,P}.
The polarization of the CMB remains undetected, its
upper limit being $P<6\times 10^{-5}$ \cite{Lubin83}.

A primordial magnetic field depolarizes the CMB radiation on
those frequencies that experience a significant amount of
Faraday rotation around the time of decoupling.
In this paper, we have applied the analytic method developed in Ref. 
\cite{Zaldarriaga95} to estimate the depolarizing effect of a 
primordial magnetic field across the last-scattering surface,
assuming a standard ionization history.
The result is expressed by eqs. (\ref{D(F)},\ref{Dlow},\ref{Dlarge})
and is represented in Fig. 1. The CMB becomes significantly depolarized
at frequencies around and below 30 GHz $(B_*/0.01\ {\rm Gauss})^{1/2}$,
below which the degree of polarization decreases quadratically 
with frequency. $B_*$ is the value of the primordial field
at a redshift $z_*=1000$, around recombination, likely to be
$10^6$ times larger than an hypothetical cosmological magnetic field
at present times. 

The average depolarizing factor depends
only upon the parameter $F$, as defined by eq. (\ref{Fbis}),
which represents the average Faraday rotation between collisions.
We have calculated the depolarizing factor $d(F)$, as given
by equation (\ref{d(F)}), in the particular case of a 
wavevector $\vec k\parallel\vec B$. In a general case,
the factor $d$ depends upon the angle between $\vec k$
and $\vec B$. This dependence integrates away in average
quantities, after the stochastic superposition of all
Fourier modes of the density fluctuations. The value derived
here for $d$ is at most an underestimate of the 
average depolarizing effect, which would eventually
start to be significant at slightly larger frequencies. 
Our derivation also assumed that Faraday rotation
accumulates over the width of the last scattering surface.
If the primordial magnetic field is very entangled over that scale, 
the depolarizing effect starts to be significant at smaller
frequencies. 

The depolarizing mechanism has a significant effect upon the anisotropy
of the CMB on small angular scales.
On those angular scales and at
frequencies such that the depolarizing effect is large, 
the damping of anisotropies by photon diffusion is reduced,
which results in a significant increase of the anisotropy
at a fixed angular scale. Besides, depolarization
reduces the contribution of the intrinsic quadrupole
anisotropy. Figure 2 displays the estimate for the 
percentual change of 
the anisotropy power spectrum at 
$l=1000$ due to the reduction in
diffusion damping, as a function
of frequency and for different values of the primordial magnetic
field at recombination.

We conclude that a primordial magnetic field increases the anisotropy  
of the CMB by up to 7.5\% at 
$l\approx 1000$ in a standard CDM cosmology.
The asymptotic strength of the effect is independent of the 
intensity of
the magnetic field, but the frequencies at which it starts to be 
significant are those around and below 30 GHz $(B_*/0.01 
{\rm Gauss})^{1/2}$.
Measurements of anisotropy and polarization at sufficiently low
frequencies could probe primordial magnetic fields in an 
interesting range. 

\section*{Acknowledgements}

DH and JH are grateful to Nathalie Deruelle and the DARC at Meudon
for hospitality while working on this project. 
The work of DH was partially supported by an EEC, DG-XII Grant 
No. CT94-0004, and by CONICET.
JH is grateful to the Royal Society for a grant.

\newpage

\begin{figure}
\begin{center}
\leavevmode
\epsfysize=4.5in
\epsfbox{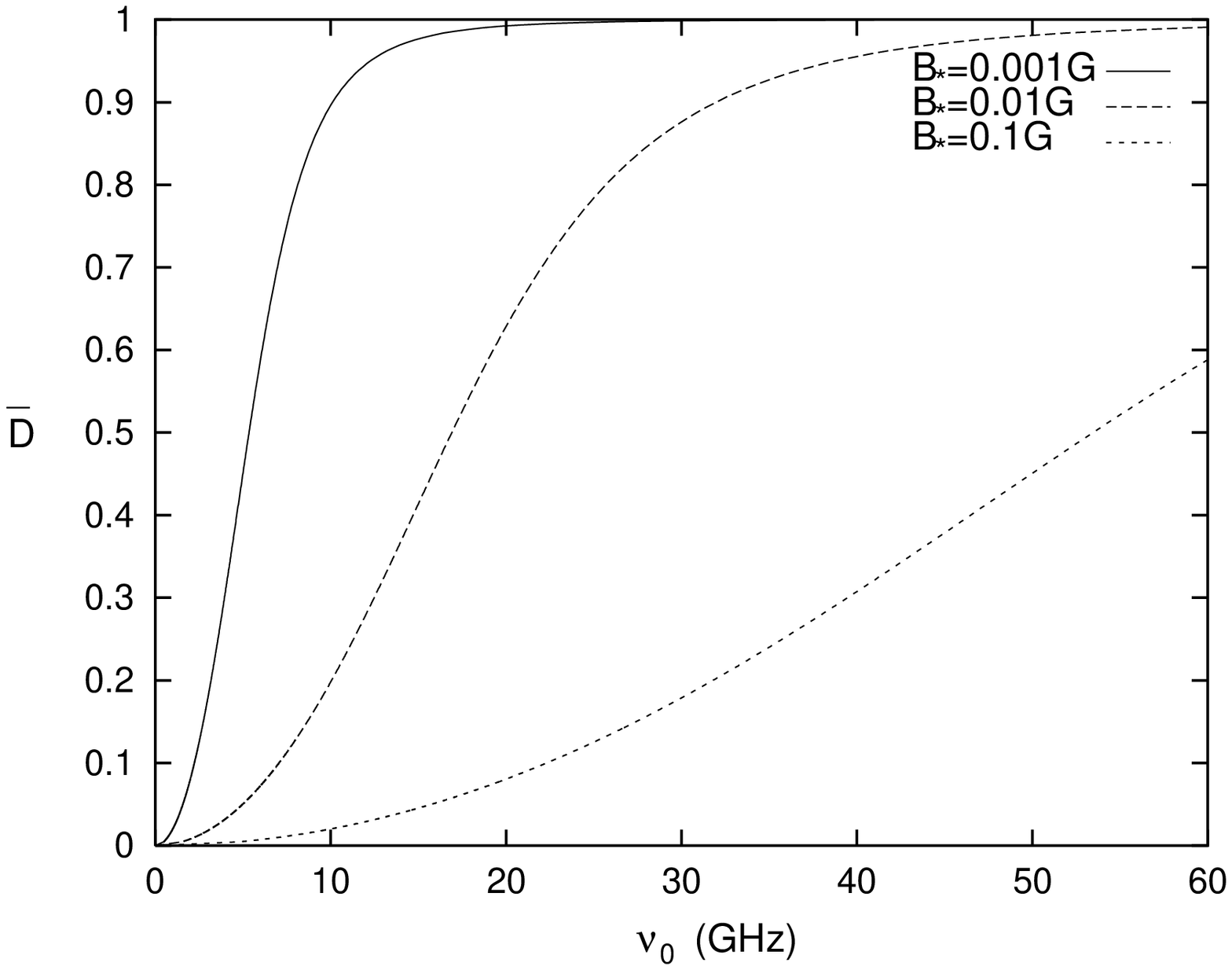}
\end{center}
\caption
{The average depolarizing factor $\bar D$ as a function of the CMB
fre\-quen\-cy $\nu_0$. The corresponding figure for an arbitrary value of
$B_*$ is identical to any of these after  a scaling of
the frequency units, proportional to $B_*^{1/2}$.}
\end{figure}

\newpage

\begin{figure}
\begin{center}
\leavevmode
\epsfysize=4.5in
\epsfbox{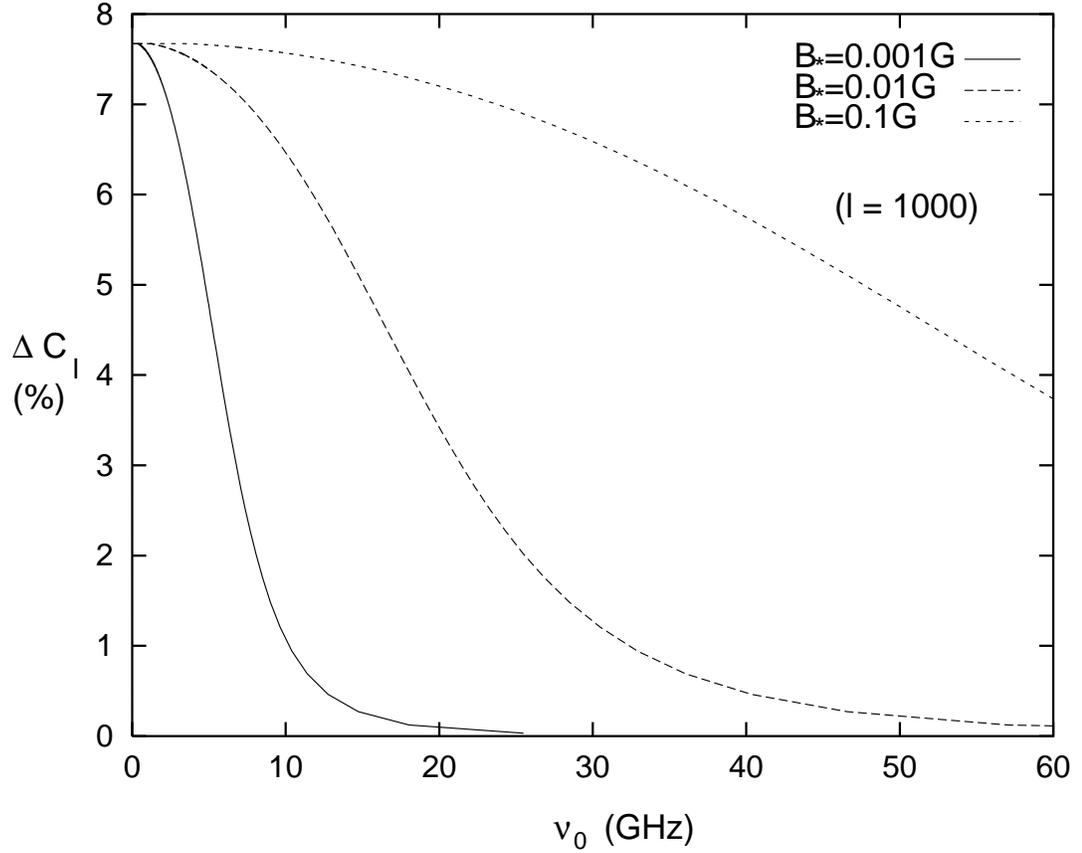}
\end{center}
\caption
{Analytic estimate of the percentual change 
due to reduction in diffusion damping
of the $l=1000$ anisotropy correlation
function multipoles as a function of the CMB frequency
for different strengths of the primordial magnetic field
at recombination.  The corresponding figure for arbitrary $B_*$
can be obtained from any of these after a scaling of the frequency 
units, proportional to $B_*^{1/2}$}
\end{figure}

\newpage

\begin{figure}
\begin{center}
\leavevmode
\epsfysize=4.5in
\epsfbox{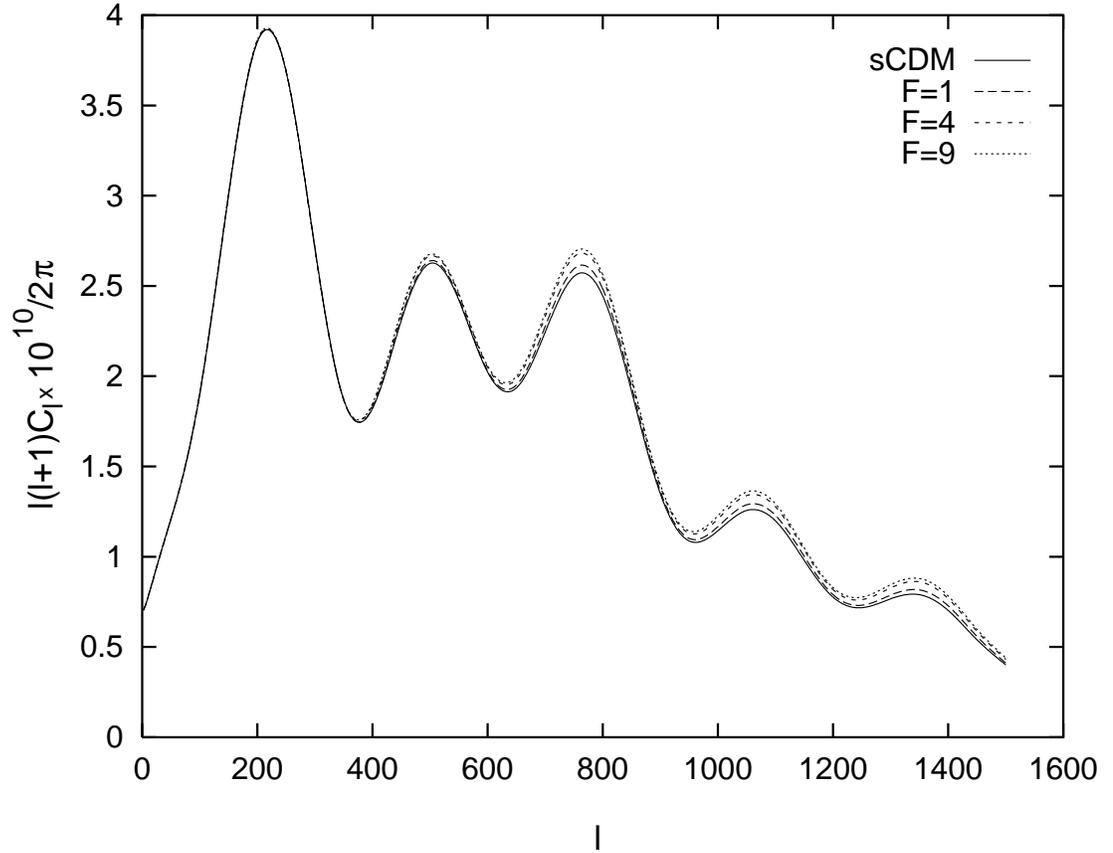}
\end{center}
\caption
{Numerical integration for the
multipoles of the anisotropy correlation function
in a standard CDM model without a primordial
magnetic field $(F=0)$, and with $F=1,\ 4,\ 9$, 
which correspond
to $\nu_0=\nu_{\rm d},\ \nu_{\rm d}/2,\ \nu_{\rm d}/3$
respectively, with $\nu_{\rm d}\approx$ 27 GHz 
$(B_*/0.01{\rm Gauss})^{1/2}$.}
\end{figure}

\newpage

\begin{figure}
\begin{center}
\leavevmode
\epsfysize=4.5in
\epsfbox{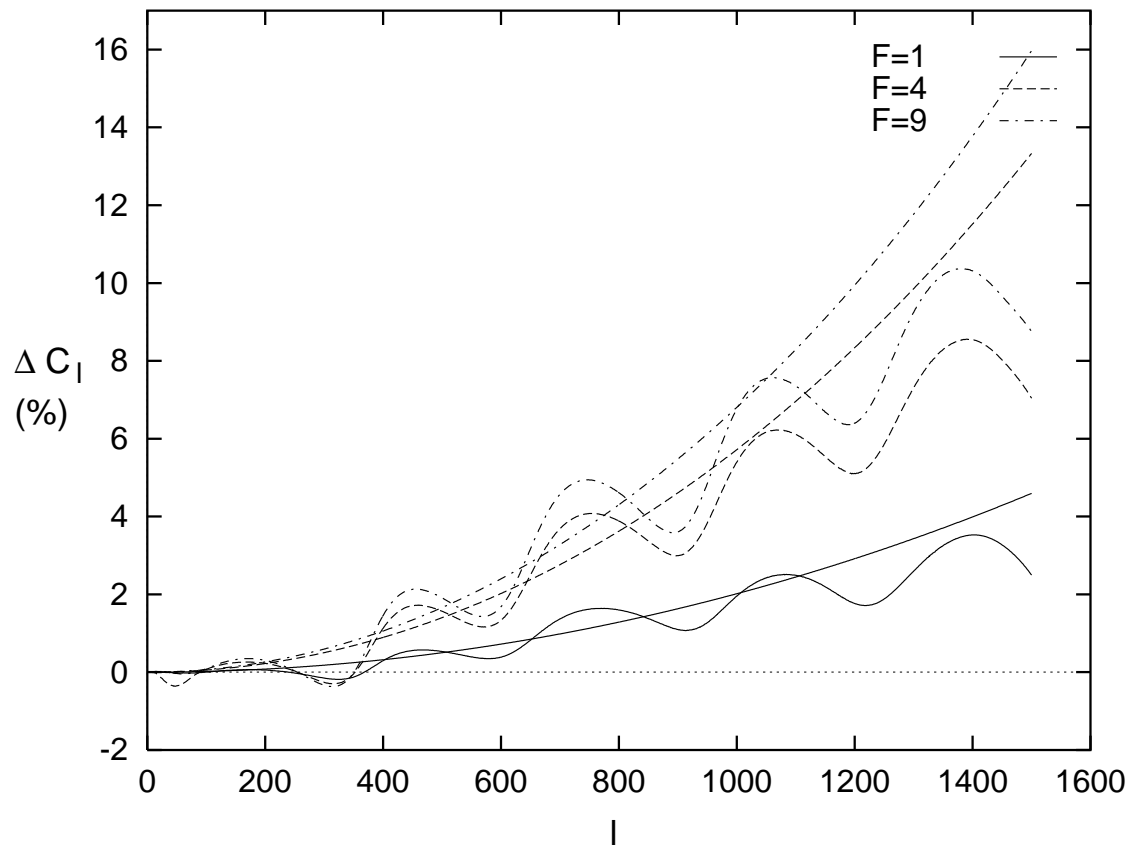}
\end{center}
\caption
{Numerical result for the percentual change of $C_l$
as a function of $l$ 
relative to its value without magnetic field 
in a standard CDM model.
The monotonic curves also shown for comparison purposes
correspond to the 
analytic estimate of the effect of reduced diffusion damping.}

\end{figure}

\end{document}